\documentclass[aps,preprint,epsfig,rotate]{revtex4}
\usepackage{graphicx}
\usepackage{bm}
\usepackage{epsfig}

\begin{document}

\title{Differential equation for the Uehling potential} 

\author{Alexei M. Frolov}
 \email[E--mail address: ]{alex1975frol@gmail.com}  

\affiliation{Department of Applied Mathematics \\
 University of Western Ontario, London, Ontario N6H 5B7, Canada}

\date{\today}

\begin{abstract}

The second-order differential equation for the Uehling potential is derived explicitly. The right side 
of this differential equation is a linear combination of the two Macdonald's functions $K_{0}(b r)$ and 
$K_{1}(b r)$. This central potential is of great interest in many QED problems, since it describes the 
lowest-order correction for vacuum polarization in few- and many-electron atoms, ions, muonic and 
bi-muonic atoms/ions as well as in other similar systems. \\  

\noindent 
PACS number(s): 12.20.-m , 12.20.Ds and 31.30.Jv \\
\end{abstract}

\maketitle

\newpage
\section{Introduction}

The main goal of this short communication is to derive the explicit second-order differential equation for 
the Uehling potential \cite{Uehl}. This central potential was introduced by Edwin Albrecht Uehling in 1935 
\cite{Uehl} when he investigated the effect of vacuum polarization which is produced by a point electric 
charge. It plays an important role in modern Quantum Electrodynamics and its application to various 
problems known in atomic, muon-atomic and nuclear physics. As is well known (see, e.g., \cite{AB}, 
\cite{BLP} and \cite{Grein}) in the lowest-order approximation the vacuum polarization, which always arises 
around an arbitrary (point) electrical charge, is described by the Uehling potential \cite{Uehl}. For atomic 
and Coulomb few- and many-body systems the Uehling potential \cite{Uehl} generates a small correction on 
vacuum polarization which must be added to the leading contribution from the original Coulomb potential. The 
sum of the original Coulomb and Uehling potentials is represent in the following integral form \cite{Grein} 
\begin{eqnarray}    
 V(r) = \frac{Q e}{r} + U(r) = \frac{Q e}{r} \Bigl[ 1 + \frac{2 \alpha}{3 \pi} \int_{1}^{\infty} 
 d\xi \Bigl( 1 + \frac{1}{2 \xi^{2}} \Bigr) \frac{\sqrt{\xi^{2} - 1}}{\xi^{2}} 
 \exp(- 2 m_e r \xi) \Bigr] \; \; \label{first} 
\end{eqnarray}
where $Q e$ is the electric charge of the central atomic nucleus, $e$ is the electric charge of the positron  
(or anti-electron), i.e., $e > 0$, $\alpha = \frac{e^{2}}{\hbar c} \approx \frac{1}{137}$ is the dimensionless 
fine-structure constant (see below), while $m_e$ is the electron mass at rest and $r$ is the electron-nuclear 
distance. In this formula and everywhere in this study (unless otherwise specified) we apply the relativistic 
units, where $\hbar = 1$ and $c = 1$. In these units the Uehling potential $U(r) (\equiv U_{ehl}(r))$ from 
Eq.(\ref{first}) is written in the form
\begin{eqnarray}    
 U(r) &=& \frac{2 Q e \alpha}{3 \pi r} \int_{1}^{\infty} \Bigl( 1 + \frac{1}{2 \xi^{2}} \Bigr) 
 \frac{\sqrt{\xi^{2} - 1}}{\xi^{2}} \exp(- 2 m_e r \xi) d\xi \; \; \nonumber \\ 
 &=& \frac{A}{r} \int_{1}^{\infty} \Bigl( 1 + \frac{1}{2 \xi^{2}} \Bigr) 
 \frac{\sqrt{\xi^{2} - 1}}{\xi^{2}} \exp(- 2 y r \xi) d\xi \; , \; \label{U1X}
\end{eqnarray} 
where $A = \frac{2 Q e \alpha}{3 \pi}$ and $y = m_{e}$. Unless otherwise is specified, everywhere 
below in this study we shall use the relativistic units, where $\hbar = 1, c = 1$. It is clear that 
the factor $y \; r$ in the exponent must be dimensionless. Therefore, the multiplier $y$ must have 
dimension $(length)^{-1}$. In relativistic units and for electronic (or atomic) systems the factor 
$y$ can only be written in the form $y = \frac{m_{e} c}{\hbar} = \frac{1}{\Lambda_{e}}$, where 
$\Lambda_{e} = \alpha^{-1} a_{0} \approx 3.8615926796 \cdot 10^{-11}$ $cm$ is the reduced electron's 
Compton wavelength, while $a_{0}$ is the Bohr (atomic) radius. As directly follows from here in 
relativistic units we have $y = 1$, while in atomic units for the dimensionless product $y \; r$ one 
finds $y \; r = \Bigl(\frac{1}{\Lambda_{e}}\Bigr) \; r = \Bigl(\frac{a_{0}}{\Lambda_{e}}\Bigr) \; 
\Bigl(\frac{r}{a_{0}}\Bigr) = \alpha^{-1} \; \Bigl(\frac{r}{a_{0}}\Bigr) = \alpha^{-1} \; r$. In 
other words, in atomic units, where $\hbar = 1, m_e = 1$ and $e = 1$, the dimensionless product 
$y \; r$ in Eq.(\ref{U1X}) equals $\alpha^{-1} \; r$, since in these units we have $a_{0} = 1$.    
 
The central potential $U(r)$ describes the lowest-order correction on vacuum polarization in few- and 
many-electron atoms, ions and muonic atoms. Applications of the Uehling potential to various atomic 
and muon-atomic systems can be found, e.g., in \cite{Dubl}, \cite{Plum} and \cite{Fro3} (see also the 
references mentioned in these papers). This potential is also of great interest in many other problems 
where one needs to evaluate the lowest-order vacuum polarization correction to the cross sections of a 
number of QED processes, including the Mott electron scattering, bremsstrahlung, creation and/or 
annihilation of the $(e^{-}, e^{+})-$pair in the Coulomb field generated by a heavy atomic nucleus. 
Taking into account the importance of the Uehling potential in a large number of QED problems we have 
decided to derive the explicit differential equation which uniformly produces (as its solution) the 
Uehling potential. This differential equation allows us to understand a very close relation between 
the Uehling potential and modified Bessel functions of the second kind $K_{n}(z)$, which are also 
called the Macdonald's functions. In turn, by using this relation one can easily predict some new 
properties of the Uehling potential $U(r)$, Eq.(\ref{U1X}).   

\section{Analytical formula for the Uehling potential}

Note that the formula, Eq.(\ref{U1X}), can also be written in the form of a finite sum \cite{FroW2012} 
(see also \cite{Fro2004}, \cite{FroCanJP} and Appendix): 
\begin{eqnarray}    
 U(r) = \frac{2 Q e \alpha}{3 \pi r} \Bigl[ \Bigl(1 + \frac{y^{2}}{3}\Bigr) \; K_{0}(2 y r) 
  - \frac{y}{6} \; Ki_{1}(2 y r) - \Bigl(\frac56 + \frac{y^{2}}{3}\Bigr) \; Ki_{2}(2 y r)\Bigr] 
  \; \; , \; \; \label{Uamf0} 
\end{eqnarray}  
which contains only the modified Bessel $K_{0}(2 y r) (= Ki_{0}(2 y r))$ function of the second order, or  
Macdonald's function(s) \cite{Mac}, \cite{Watson}. The notation $Ki_{n}(z)$ (where $n = 0, 1, 2, \ldots$) 
stands for the successive (or multiple) integrals of this function defined exactly as in \cite{AS}, i.e., 
the $Ki_{1}(z)$ and $Ki_{2}(z)$ functions are 
\begin{eqnarray}    
 Ki_{1}(z) = \int_{z}^{\infty} K_{0}(x) dx \; \; \; {\rm and} \; \; \;  Ki_{2}(z) = \int_{z}^{\infty} 
 Ki_{1}(x) dx \; \; . \; \; \label{Uamf1} 
\end{eqnarray} 

The formula for the Uehling potential can also be written in other different (but equivalent!) forms. For 
instance, by using the known relations \cite{AS} between the $Ki_{n}(z)$ functions (or integrals) we can 
write another formula for the $U(r)$ potential  
\begin{eqnarray}    
 U(r) = \frac{2 Q e \alpha}{3 \pi r} \Bigl[ K_{0}(2 y r) - \frac12 \; Ki_{2}(2 y r) - \frac12 \; 
 Ki_{4}(2 y r) \Bigr] \; \; . \; \; \label{Uamf2} 
\end{eqnarray} 
However, in this study we restrict ourselves to the formula, Eq.(\ref{Uamf0}), only. In order to simplify 
our calculations below let us introduce the new parameter $b = 2 y$ in Eq.(\ref{Uamf0}), which takes the 
form 
\begin{eqnarray}    
 U(r) = \frac{2 Q e \alpha}{3 \pi r} \Bigl[ \Bigl(1 + \frac{b^{2}}{12}\Bigr) \; K_{0}(b r) - \frac{b}{12} 
 \; Ki_{1}(b r) - \Bigl(\frac56 + \frac{b^{2}}{12}\Bigr) \; Ki_{2}(b r)\Bigr] \; \; , \; \; \label{Uamf2} 
\end{eqnarray} 
or
\begin{eqnarray}    
 U(r) = \frac{Q e \alpha}{18 \pi r} \Bigl[ \Bigl(12 + b^{2}\Bigr) \; K_{0}(b r) - b \; Ki_{1}(b r) - 
 \Bigl(10 + b^{2}\Bigr) \; Ki_{2}(b r)\Bigr] \; \; , \; \; \label{Uamf3} 
\end{eqnarray} 
This explicit expression for the Uehling potential is appropriate and sufficient for our following 
transformations required in this study (see below). 

\section{Derivation of the differential equation for the Uehling potential}

In this Section we derive the second-order differential equation for the Uehling potential, i.e., this 
Section is the central part of our study. To achieve this goal let us re-write the Uehling potential 
$U(r) \equiv U_{ehl}(r)$ in the form (see, e.g., \cite{FroCanJP}):  
\begin{eqnarray}    
 U(r) = \frac{A}{r} \Bigl[ \Bigl(12 + b^{2}\Bigr) \; K_{0}(b r) - b \; Ki_{1}(b r) - \Bigl(10 + 
  b^{2}\Bigr) \; Ki_{2}(b r) \Bigr] \; \; , \; \; \label{U1} 
\end{eqnarray} 
where in the relativistic units $A = \frac{\alpha Q e}{18 \pi}$ and $b = 2 m_{e}$. From Eq.(\ref{U1}) 
one easily finds 
\begin{eqnarray}    
 [r U(r)] = A \Bigl[ \Bigl(12 + b^{2}\Bigr) \; K_{0}(b r) - b \; Ki_{1}(b r) - \Bigl(10 + 
  b^{2}\Bigr) \; Ki_{2}(b r) \Bigr] \; \; . \; \; \label{U11} 
\end{eqnarray} 
The first-order derivative from this expression in respect to the $r-$variable is 
\begin{eqnarray}    
 \frac{d [r U(r)]}{d r} = [r U(r)]^{\prime} = A \Bigl[ - \Bigl(12 b + b^{3}\Bigr) \; K_{1}(b r) 
  + b^{2} \; K_{0}(b r) + \Bigl(10 b + b^{3}\Bigr) \; Ki_{1}(b r) \Bigr]    \; \; . \; \; 
  \label{U12} 
\end{eqnarray} 
To obtain this formula we have used the following, well known relations (see, e.g., \cite{AS} and 
\cite{GR}) for the modified Bessel functions 
\begin{eqnarray}    
 \frac{d K_{0}(b r)}{d r} = (- b) K_{1}(b r) \; \; , \; \; \frac{d Ki_{1}(b r)}{d r} = (- b) 
  K_{0}(b r) \; \; {\rm and} \; \;  \frac{d Ki_{2}(b r)}{d r} = (- b) Ki_{1}(b r) \; \; . \; \; 
  \label{U13} 
\end{eqnarray} 

From the formula, Eq.(\ref{U12}), we can determine the second-order derivative in respect to the 
$r-$variable
\begin{eqnarray}    
 & &\frac{d^{2} [r U(r)]}{d r^{2}} = [r U(r)]^{\prime\prime} = A \Bigl[ \Bigl(12 b + b^{3} \Bigr) 
  \; \Bigl( b K_{0}(b r) + \frac{1}{r} K_{1}(b r) \Bigr) - b^{3} \; K_{1}(b r) \nonumber \\
  &-& \Bigl(10 b^{2} + b^{4}\Bigr) \; Ki_{0}(b r) \Bigr] \nonumber \\
  &=& A \Bigl\{ \Bigl( 12 b^{2} + b^{4} \Bigr) \; K_{0}(b r) 
  + \Bigl[ \frac{1}{r} \Bigl( 12 b + b^{3} \Bigr) - b^{3} \Bigr] \; K_{1}(b r) 
 - \Bigl( 10 b^{2} + b^{4} \Bigr) \; K_{0}(b r) \Bigr\} \; \; . \; \; \label{U14} 
\end{eqnarray} 
To derive this formula we used two (of three) equations mentioned in Eq.(\ref{U13}) and one 
additional, well known relation (see, e.g., \cite{GR})
\begin{eqnarray}  
 \frac{d K_{\nu}(z)}{d z} = - \Bigl(\frac{\nu}{z}\Bigr) K_{\nu}(z) - K_{\nu-1}(z) \; \; , \; \; 
 \label{U133} 
\end{eqnarray} 
or
\begin{eqnarray} 
 \frac{d K_{\nu}(b r)}{d r} = b \frac{d K_{\nu}(x)}{d x} = - b \Bigl(\frac{\nu}{b r}\Bigr) 
 K_{\nu}(b r) - b K_{\nu-1}(b r) = - \Bigl(\frac{\nu}{r}\Bigr) K_{\nu}(b r) -  b K_{\nu-1}(b r) 
 \; \; , \; \; \label{U135}
\end{eqnarray}  
where in our case $\nu = 1$. The final formula for the second-order derivative $\frac{d^{2} 
[r U(r)]}{d r^{2}}$ takes the form 
\begin{eqnarray}    
 \frac{d^{2} [r U(r)]}{d r^{2}} = [r U(r)]^{\prime\prime} = A \; b \; \Bigl\{ 2 \; b \; K_{0}(b r) + 
  \frac{1}{r} \Bigl[ 12 + b^{2} (1 - r) \Bigr] \; K_{1}(b r) \Bigr\} \; \; . \; \; \label{U15} 
\end{eqnarray} 
It is clear that the last formula can also be re-written in the form 
\begin{eqnarray}    
 r \; \frac{d^{2} [U_{ehl}(r)]}{d r^{2}} + 2 \frac{d U_{ehl}(r)}{d r} =  A \; b \; \Bigl\{ 2 \; b 
  \; K_{0}(b r) + \frac{1}{r} \Bigl[ 12 + b^{2} (1 - r) \Bigr] \; K_{1}(b r) \Bigr\} \; \; , \; \; 
  \label{U155} 
\end{eqnarray} 
where we have designated the Uehling potential by the notation $U_{ehl}(r)$ which is identical to 
the potential $U(r)$ used in the equations above. 

The formula, Eq.(\ref{U155}), finally solves the problem, which has been formulated in the Introduction 
as the main goal for this our study. Indeed, the formula, Eq.(\ref{U155}), (also the formula, 
Eq.(\ref{U15})) is the required differential equation of the second-order for the Uehling potential. 
This differential equation uniformly determines the Uehling potential. Note that in respect to the 
general theory of ordinary differential equations (see, e.g., \cite{Ince}) such an uniform restoration 
of the solution of any differential equation of the second-order is possible, if (and only if) we know 
the two `initial conditions' for the unknown function and its first-order derivative. However, in this 
study we consider a different problem. Indeed, we derive (or restore) a differential equation for the 
Uehling potential from its integral representation, Eq.(\ref{U1X}), which is know $a$ $priory$. This 
means that we can always determine the numerical value of the Uehling potential $U(r)$ at any radial 
point $r = r_{a}$ by using the original formula Eq.(\ref{U1X}) and its first-order radial derivative 
$\frac{d U(r)}{d r}$. For such a derivative we have to apply the formula 
\begin{eqnarray}    
 \frac{d U(r)}{d r} = - \frac{4 Q e \alpha m_e}{3 \pi r} \int_{1}^{\infty} \Bigl( 1 + \frac{1}{2 
  \xi^{2}} \Bigr) \frac{\sqrt{\xi^{2} - 1}}{\xi} \exp(- 2 m_e r \xi) d\xi \; \; , \; \; \label{U1Y} 
\end{eqnarray} 
which directly follows from Eq.(\ref{U1X}). By performing accurate numerical integration in the 
both formulas, Eqs.(\ref{U1X}) and (\ref{U1Y}), one easily obtains the correct `initial 
conditions' at any radial point $r_a$.

To conclude this Section we have to make a remark about the units used in physical problems related 
with the lowest-order correction on vacuum polarization, or in other words, in all problems where 
the Uehling potential appears explicitly. This remark is mainly important for mathematicians and 
for all newcomers in the area of vacuum polarization. As it is mentioned above all formulas for the 
Uehling potential are written in the relativistic units, where $\hbar = 1$ and $c = 1$. However, in 
a large number of applications the same formulas are needed in atomic units, where $\hbar = 1, m_{e} 
= 1$ and $e = 1$. Therefore, it is important to know the numerical factors which are used to 
re-calculate some fundamental physical values such as mass, length, time and energy from relativistic 
to atomic units. The general philosophy and basic technique of such a process is well discussed in 
\cite{Mandl}. By using the method from \cite{Mandl} one easily obtains the following formula, 
Eq.(\ref{U1Z}), for the Uehling interaction energy in atomic units for the electron-nucleus 
interaction. Such an interaction is the product of the electric charge of electron $(- e)$ and 
Uehling potential $U(r) (\equiv U_{ehl}(r))$, Eq.(\ref{U1X}). The final formula (in atomic units) 
takes the form 
\begin{eqnarray}    
 E_{Uehl} = - e U(r) = \frac{2 Q \alpha^{2}}{3 \pi r} \int_{1}^{\infty} \Bigl( 1 + \frac{1}{2 
  \xi^{2}} \Bigr) \frac{\sqrt{\xi^{2} - 1}}{\xi^{2}} \exp(- 2 \alpha^{-1} r \xi) d\xi \; , \; 
  \label{U1Z} 
\end{eqnarray} 
where in atomic units the parameter $y$ in Eq.(\ref{U1X}) equals: $y = \alpha^{-1}$ (see above). Note 
that the numerical value of speed of light in vacuum (in atomic unis) also equals $c = \alpha^{-1}$ 
exactly, where $\alpha \approx 7.2973525693 \cdot 10^{-3}$ \cite{NIST} is the dimensionless 
fine-structure constant. 

\section{Conclusion}

We have derived the differential equation of the second-order for the Uehling potential (see, 
Eqs.(\ref{U15}) and (\ref{U155})). Our derivation is absolutely transparent, explicit (all intermediate 
steps are shown in detail) and simple. The arising second-order differential equation is also relatively 
simple and this fact drastically simplifies investigations of the newly derived equation and its 
analytical and/or numerical solutions. In general, by using the well known properties of Macdonald's 
functions (see, e.g., \cite{Watson}) one can re-write the both our equations, Eqs.(\ref{U15}) - 
(\ref{U155})), into a number of different (but equivalent!) forms. 

Note that the right side of our differential equation, Eqs.(\ref{U155}) (see also Eq.(\ref{U15})), 
contains only the two modified Bessel functions of the second kind (or Macdonald's functions) $K_{0}(b 
r)$ and $K_{1}(b r)$. This indicates clearly that there is a very close relation between the Uehling 
potential and Macdonald's functions $K_{n}(z)$ of the lowest orders and multiple successive (or 
multiple) integrals of these functions. Based on this relation we can investigate and discover some new 
properties of the Uehling potential.   

\appendix
\section{On the finite analytical formula for the Uehling potential}
\label{A} 

Let us briefly discuss the derivation of the finite analytical formula for the Uehling potential. This 
problem has been considered earlier in \cite{FroW2012}, \cite{Fro2004} and \cite{FroCanJP}. The original 
problem was complicated by wrong statements in dozens of modern QED books and textbooks (see, e.g., 
\cite{AB} - \cite{Grein}) that the `closed analytical expression for the Uehling potential does not 
exist' and/or `it cannot be derived in a closed form'. In order to illustrate the fallacy of such 
statements let us directly obtain the closed analytical expression for the following integral $I(r)$
\begin{eqnarray}  
 I(r) = r \; U(r) = \int_{1}^{\infty} \Bigl( 1 + \frac{1}{2 \xi^{2}} \Bigr) \frac{\sqrt{\xi^{2} - 
 1}}{\xi^{2}} \exp(- 2 m r \xi) d\xi \label{A1}  
\end{eqnarray}  
which essentially (up to the factor $A = \frac{2 Q e \alpha}{3 \pi}$) coincides with the product of the 
Uehling potential $U(r)$ and variable $r$. At the first step we introduce the new variable $zeta$, where 
$\cosh \zeta = \xi$. In this new variable we can write $d\xi = \sinh\zeta d\zeta, \sqrt{\xi^{2} - 1} = 
\sinh\zeta$ and $\sinh^{2}\zeta = \cosh^{2}\zeta - 1$. By using these formulas one transforms the 
integral, Eq.(\ref{A1}), to the form 
\begin{eqnarray}  
 I(r) &=& \int_{0}^{\infty} \Bigl( 1 + \frac{1}{2 \cosh^{2}\zeta} \Bigr) \; 
 \Bigl(\frac{\sinh^{2}\zeta}{\cosh^{2}\zeta}\Bigr) \; \exp(- 2 m r \cosh \zeta) \; d\zeta \nonumber \\ 
 &=& \int_{0}^{\infty} \Bigl( 1 - \frac{1}{2 \cosh^{2}\zeta} - \frac{1}{2 \cosh^{4}\zeta}\Bigr) \; 
 \exp(- 2 m r \cosh \zeta) \; d\zeta \\ 
 &=& Ki_{0}\Bigl(\frac{2 r}{\Lambda_{m}}\Bigr) - \frac12 Ki_{2}\Bigl(\frac{2 r}{\Lambda_{m}}\Bigr) 
 - \frac12 Ki_{4}\Bigl(\frac{2 r}{\Lambda_{m}}\Bigr) \label{A2}  
\end{eqnarray} 
where $Ki_{0}(z) \equiv K_{0}(z)$ is the Macdonald's function of zero order \cite{Mac}, $Ki_{n}$ are the 
successive (or multiple) integrals of this function defined exactly as in \cite{AS} (see also 
Eq.(\ref{Uamf1}) in the main text). The notation $\Lambda_{m} = \frac{\hbar}{m c}$ stands for the Compton 
wavelength for the particle with mass $m$. Finally, our formula for the integral, Eq.(\ref{A2}), 
essentially coincides with Eq.(\ref{Uamf2}) from the main text.     

At the second step of this procedure we have to apply (twice) the formula, Eq.(11.2.14), from \cite{AS} 
\begin{eqnarray} 
  n Ki_{n+1}(z) = - z Ki_{n}(z) + (n - 1) Ki_{n-1}(z) + z Ki_{n-2}(z) \; . \; \label{A3}
\end{eqnarray} 
In the first case we apply this formula for $n = 3$, while in the second case we have to assume that $n 
= 2$. This allows us to reduce the formula, Eq.(\ref{A2}), to its final form 
\begin{eqnarray}    
 I(r) = r \; U(r) = \Bigl(1 + \frac{z^{2}}{12}\Bigr) \; K_{0}(z) - \frac{z}{12} \; Ki_{1}(z) - 
 \Bigl(\frac56 + \frac{z^{2}}{12}\Bigr) \; Ki_{2}(z) \; \; , \; \; \label{A4} 
\end{eqnarray}
where $z = \frac{2 r}{\Lambda_{m}}$. This expression essentially coincides with the formula, 
Eq.(\ref{Uamf2}), from the main text. An obvious advantage of this formula follows from the fact that this 
formula contains only the lowest-order $K_{0}(z)$ Macdonald's function and two lowest-order multiple 
integrals of this function, i.e., the $Ki_{1}(z)$ and $Ki_{2}(z)$ functions. Disadvantage of this formula 
is also clear, since the coefficients in the front of all three $K_{0}(z), Ki_{1}(z)$ and $Ki_{2}(z)$ 
functions in this equation are now $z-$dependent (or $r-$dependent).


\begin{thebibliography}{99}

\bibitem{Uehl} E.A. Uehling, Phys. Rev. {\bf 48}, 55 (1935). 

\bibitem{AB} A. Akhiezer and V.B. Berestetskii, \textit{Quantum Electrodynamics} (4th ed., Science, Moscow 
(1981)) [in Russian]. 

\bibitem{BLP} V.B. Berestetskii, E.M. Lifshitz and L.P. Pitaevskii, \textit{Relativistic Quantum Theory} 
(Pergamon Press, Oxford (1971)). 

\bibitem{Grein} W. Greiner and J. Reinhardt, \textit{Quantum Electrodynamics} (4th ed., Springer Verlag, 
Berlin (2009)).

\bibitem{Dubl} T. Dubler, K. Kaeser, B. Robert-Tissot, L.A. Schaller, L. Schellenberg and H. Schneuwly, 
Nucl. Phys. A {\bf 294}, 397 (1978).  

\bibitem{Plum} G. Plunien and G. Soff, Phys. Rev. A {\bf 51}, 1119 (1995).  

\bibitem{Fro3} A.M. Frolov, J. Comput. Science {\bf 5}, 499 (2014). 


\bibitem{FroW2012} A.M. Frolov and D.M. Wardlaw, Eur. Phys. J. B (2012) (Solid State and
Complex Systems) {\bf 85}, 348 (2012). 

\bibitem{Fro2004} A.M. Frolov, J. Phys. B {\bf 37}, 4517 (2004).

\bibitem{FroCanJP} A.M. Frolov, Can. J. Phys. {\bf 92}, 1094 (2014). 

\bibitem{Mac} H.M. Macdonald, Proc. London Math. Soc. {\bf XXX}, 167 (1899). 

\bibitem{Watson} G.N. Watson, \textit{a treatise on the THEORY OF BESSEL FUNCTIONS}. 
(2nd edition, Cambridge University Press, London (1944), reprinted in 1966). 

\bibitem{AS} \textit{Handbook of Mathematical Functions} (M. Abramowitz and I.A. Stegun (Eds.), 
Dover Publ. Inc., New York (1972)). 

\bibitem{GR} I.S. Gradstein and I.M. Ryzhik, \textit{Tables of Integrals, Series and Products} 
(6th revised ed., Academic Press, New York (2000)).

\bibitem{Mandl} F. Mandl and G. Shaw, \textit{Quantum Field Theory} (2nd ed., John Wiley and 
Sons, Ltd. (2010)).

\bibitem{NIST} see, e.g., https://physics.nist.gov/cgi-bin/cuu/Value?

\bibitem{Ince} E.L. Ince, \textit{Ordinary Differential Equations}  (Dover Publ. Inc., Mineola, 
New York (2012)).  

\end{thebibliography}
\end{document}